\documentclass[superscriptaddress, aps, pre, reprint, floatfix]{revtex4-2}
\usepackage{graphicx,hyperref,color,upgreek}
\usepackage{amsmath}
\usepackage{latexsym}
\usepackage{float}
\usepackage{amssymb}
\usepackage{multirow}
\usepackage{graphicx}
\usepackage{textcomp}
\usepackage{hyperref}
\usepackage{array}
\usepackage{xcolor}
\usepackage{hyperref}
\usepackage{xkeyval,xcolor}
\usepackage{xcite}
\graphicspath{{Main/figures/},{Main/}}
\usepackage[caption=false]{subfig}
\usepackage{dcolumn}
\usepackage{bm}
\usepackage{xr}
\usepackage{tikz}
\usepackage{tabularx}
\makeatletter
\newcommand*{\addFileDependency}[1]{
  \typeout{(#1)}
  \@addtofilelist{#1}
\IfFileExists{#1}{}{\typeout{No file #1.}}
}

\makeatother

\begin{document}


\title{Collective Behavior of Intelligent Active Brownian Particles in the Presence of a Static Obstacle}
\author{Surya Narayan Sahoo}
\affiliation{
Department of Physics, School of Basic Sciences,
Indian Institute of Technology Bhubaneswar,
Jatni, Odisha-752050, India
}
\author{Arabinda Bera}
\affiliation{
Institute of Physics,
University of Greifswald,
Felix-Hausdorff-Stra{\ss}e 6,
17489 Greifswald, Germany
}
\author{Jiarul Midya}
\email{jmidya@iitbbs.ac.in}
\affiliation{
Department of Physics, School of Basic Sciences,
Indian Institute of Technology Bhubaneswar,
Jatni, Odisha-752050, India
}
\date{\today}
\begin{abstract}
Using computer simulations, we investigate the collective behavior of two-dimensional active Brownian particles (ABPs) with excluded-volume interactions in the bulk and in the presence of a single static circular obstacle. Perception-mediated interactions are introduced through a vision-based steering mechanism that enables each particle to reorient its propulsion direction according to the instantaneous positions of neighboring particles within a prescribed vision cone. In the bulk, these intelligent active Brownian particles (iABPs) exhibit distinct collective states, including aggregated clusters, worm-like chains, worm–aggregate coexistence, and dilute gas phases. We characterize these states using a shape anisotropy parameter and construct the corresponding phase diagrams. The presence of a static obstacle, which interacts with the particles through purely repulsive forces, qualitatively alters their collective behavior. In contrast to conventional ABPs, whose isotropic accumulation around the obstacle increases with activity, iABPs exhibit the opposite trend, with boundary accumulation decreasing as the activity increases. We further identify empirical scaling relations that describe particle accumulation and cluster formation at the obstacle boundary. Consequently, the relative effective diffusion coefficient of iABPs displays a nonmonotonic dependence on self-propulsion speed, whereas that of conventional ABPs decreases monotonically with increasing activity. In addition, we show that the residence times of iABPs are orders of magnitude shorter than those of conventional ABPs, indicating that perception-mediated interactions can be useful for controlling the organization and transport of active particles in complex environments.

\end{abstract}

\maketitle
\section{Introduction} \label{intro}
Active matter systems consist of self-driven units that continuously consume energy to generate motion and mechanical stresses, thereby producing collective dynamics far from thermodynamic equilibrium \cite{Kubo1991, Ramaswamy_2010}. A wide range of biological and synthetic systems fall within this class, including bacterial suspensions, cellular assemblies, animal groups, and artificial microswimmers, in which large-scale self-organization, often spanning several orders of magnitude larger than the size of individual units, emerges from local interactions without centralized control \cite{Vicsek_2012, AndreaAnnuRev2014, ElgetiRPP2015, BenJacob_1994, BALLERINI2008201}. A central focus in active matter research is the development of minimal models that capture such emergent phenomena, including flocking, clustering, and order--disorder transitions \cite{GuillaumePhysRevLett2004, SokolovPhysRevLett2007, Vicsek_2012, CatesARCMP2015, CapriniPRL2023, DigregorioPRL2018, SolonPhysRevLett2015}. One of the simplest and most widely studied frameworks is the active Brownian particle (ABP) model, which combines persistent self-propulsion with rotational diffusion and short-range repulsive interactions \cite{tenHagenJPCM2011, FilyPRL2012, RednerPhysRevLett2013}. At sufficiently high densities, 
crowding hinders persistent motion of individual ABPs, leading to nonequilibrium clustering and motility-induced phase separation, phenomena that have been extensively investigated both computationally \cite{RednerPhysRevLett2013, ButtinoniPhysRevLett2013, CaporussoPhysRevLett2020, DittrichPhysRevE2023, OthmanPhysRevE2025} and experimentally \cite{ TheurkauffPhysRevLett2012, Palacciscience2013, NishiguchiPhysRevE2015, GeyerPhysRevX2019}. This demonstrates that activity alone can drive macroscopic organization even in the absence of explicit alignment interactions.

Many active agents, however, can sense their environment and adjust their motion accordingly \cite{DanielPNAS2014, BaeuerleNatCommun2018, SinghPhysRevE2020, ZhaoPhysRevLett2023}. A prominent example is bird flocks, where individuals rely on visual perception to interact with neighbors and avoid collisions during navigation \cite{BaeuerleNatCommun2018}, leading to large-scale coordinated motion. To incorporate perception-driven effects into minimal active-matter frameworks, models such as the Vicsek model \cite{TVicsekPhysRevLett1995} and its variants \cite{GregoirePhysRevLett2004, MartinSoftMater2018, KurstenPhysRevLett2020, ZhaoPhysRevE2021} have been proposed, in which such interactions are captured in a coarse-grained manner via local velocity alignment between neighboring particles. Recently, a model of intelligent active Brownian particles (iABPs) has been introduced, in which the propulsion direction dynamically adapts to the positions of neighboring particles within a prescribed vision cone \cite{BarberisPhysRevLett2016, RNegiSoftMatter2022}. These iABPs in bulk exhibit qualitatively richer collective behavior than conventional ABPs, including compact aggregates, elongated worm-like chains, coexistence of worm-aggregate phases, and milling states \cite{BarberisPhysRevLett2016, RNegiSoftMatter2022, RNegiPRR2024, RChandraJCP2025, RodrigoPhysRevLett2024}. The resulting phase behavior is sensitive to the vision angle, interaction strength, and activity, highlighting the crucial role of anisotropic sensing in formation of emergent structures. 

Although bulk studies of perception-mediated interactions in active particles provide important insights, most realistic environments are intrinsically heterogeneous and confined. Under these conditions, the collective behavior of active particles has attracted considerable attention due to its relevance to microbial transport \cite{BhattacharjeeNatCommun2019, MengPRR2023}, particle sorting \cite{NitinPhysRevE2019}, and the design of synthetic active materials for drug delivery \cite{DaddiJCP2019}. Experimental and theoretical studies on bacterial suspensions, algae, sperm cells, and artificial microswimmers have demonstrated that the presence of planar or curved solid obstacles can strongly influence transport and self-organization, and leads to boundary accumulation, trapping, rectification, vortex formation, and active turbulence ~\cite{ChepizhkoPRL2013, PhysRevE.97.032606, ReichhardtPhysRevE2020, IraniPRL2022, KuipouSciRep2023, MengPRR2023,  Mukherjee2023, PattanayakEPJE2019}. In particular, active particles near convex obstacles often exhibit localized aggregation and persistent orbital motion, while geometric confinement can substantially modify diffusion and clustering dynamics ~\cite{DasPhysRevE2020, Tiwari_2024, Mokhtari_2017}. Many of these nonequilibrium behaviors are well captured by minimal models of active Brownian particles (ABPs), where steric interactions alone can give rise to clustering and confinement-driven vortex states~\cite{DasPhysRevE2020, Tiwari_2024,  MozaffariPhysRevFluids2018, PhysRevE.97.032606}. However, most existing studies focus on conventional (“dumb”) active particles interacting through steric forces, while the role of perception-mediated steering in the presence of obstacles remains largely unexplored. Here, we address this gap by investigating the collective dynamics of intelligent active Brownian particles (iABPs) around single static circular obstacles and systematically comparing their behavior with that of conventional ABPs.

In this work, we investigate the collective behavior of two-dimensional intelligent active Brownian particles in bulk and in the presence of a single static circular obstacle by systematically varying control parameters, such as the vision angle, vision range, and self-propulsion velocity. Depending on these parameters, iABPs in bulk exhibit four distinct collective phases, which are characterized using a shape-anisotropy parameter to construct the corresponding phase diagrams. In the presence of a static obstacle, particles accumulate at the boundary. While conventional ABPs exhibit an increasing degree of isotropic accumulation with increasing self-propulsion speed, iABPs display inhomogeneous aggregation and a reduced degree of boundary accumulation as activity increases. We further find that the residence time of iABPs is orders of magnitude smaller than that of conventional ABPs, indicating a reduced affinity for the obstacle boundary. These findings demonstrate that visual perception provides an efficient mechanism for active particles to avoid persistent boundary accumulation, thereby enhancing transport in bulk environments.

The rest of the manuscript is organized as follows. In Sec.~\ref{sec:modelANDmethod}, we describe the minimal model incorporating visual perception and the simulation methods employed. In Sec.~\ref{sec:results}, we first present the phase behavior of iABPs in bulk, followed by their behavior in the presence of obstacles. Finally, in Sec.~\ref{sec:conclusion}, we summarize our findings and discuss the conclusions along with future outlook.

\begin{figure}[tb]
    \centering
    \includegraphics[width=1.0\linewidth]{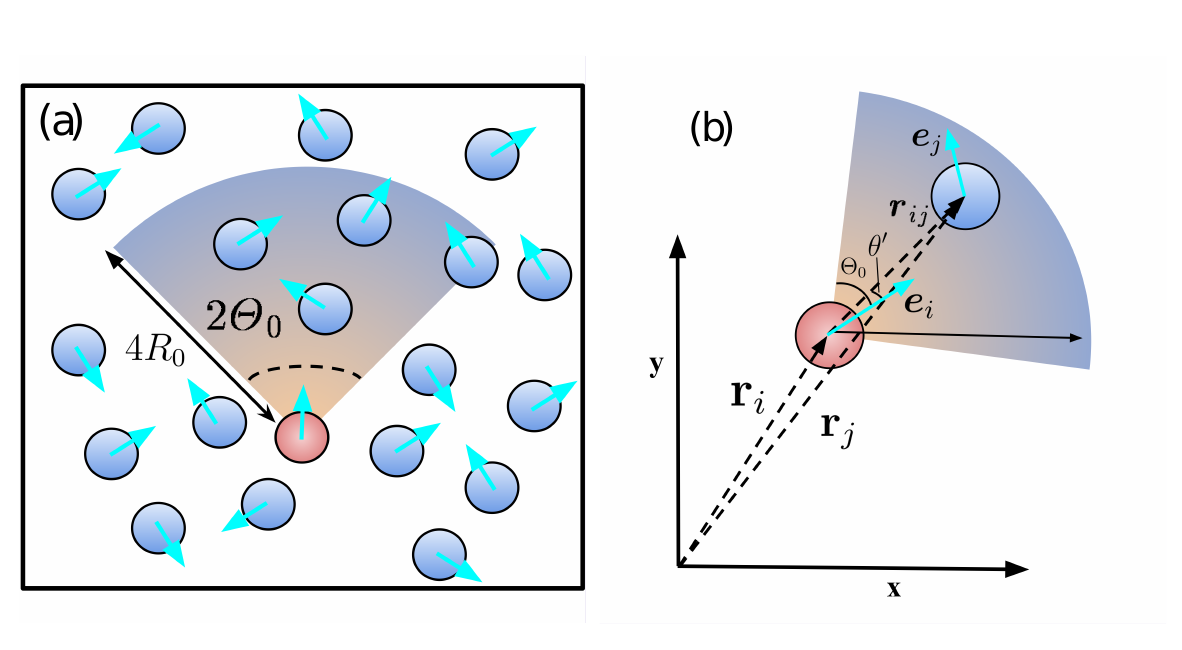}
    \caption{Schematic of the iABP model. (a) Vision cone of the $i^{\mathrm{th}}$ particle with opening angle $2\Theta_0$ centered around its propulsion direction ${\bf e}_i$ and finite vision range $R_{\mathrm{v}}=4R_0$; only particles within this sector contribute to perception. (b) A particle $j$ (position ${\bf r}_j$, orientation ${\bf e}_j$) is considered a neighbor of particle $i$ (position ${\bf r}_i$, orientation ${\bf e}_i$) if the relative angle $\theta'$ between $\hat{\bf r}_{ij} = ({\mathbf r}_j - {\mathbf r}_i)/|{\mathbf r}_j - {\mathbf r}_i|$ and ${\bf e}_i$ satisfies $\theta'<\Theta_0$ and the separation $r_{ij}<R_{\mathrm{v}}$.
    }
    \label{fig:modelSchematic}
\end{figure}
\section{Model and Method}\label{sec:modelANDmethod}
We consider a system of $N_{\rm act}$ iABPs confined in a two-dimensional square box of size $L$. The position of the $i^{\rm{th}}$ particle at time $t$ is denoted by ${\bf r}_i(t)$, with $i=1,2,\dots,N_{\rm act}$. Each particle self-propels with a constant speed $v_0$ along its instantaneous orientation ${\bf e}_i=(\cos\theta_i,\sin\theta_i)$. The translational dynamics is governed by the overdamped Langevin equation \cite{Stenhammar2014,DittrichPhysRevE2023,RNegiSoftMatter2022}
\begin{equation}
\dot{\bf r}_i = v_0{\bf e}_i + \mu {\bf F}_i^{\rm int} + \sqrt{2D_{\rm T}}\,\boldsymbol{\eta}_i(t),
\label{eq:transDyna}
\end{equation}
where $\mu=D_{\rm T}/k_{\rm B}T$ is the mobility, $D_{\rm T}$ is the translational diffusion coefficient, and ${\bf F}_i^{\rm int}$ is the total conservative force acting on particle $i$. The stochastic term $\boldsymbol{\eta}_i(t)$ represents Gaussian white noise with zero mean and unit variance, satisfying $\langle \eta_i^\alpha(t)\eta_j^\beta(t') \rangle = \delta_{\alpha\beta}\delta_{ij}\delta(t-t')$, where $\alpha,\beta \in \{x,y\}$.

The orientational dynamics of the $i$th particle is described by \cite{RNegiSoftMatter2022}
\begin{equation}
\dot{\theta}_i = \frac{\Omega}{N_{c,i}} \sum_{j\in {\rm VC}} e^{-r_{ij}/R_0} \sin(\phi_{ij}-\theta_i) + \sqrt{2D_{\rm R}}\,{\zeta}_i(t),
\label{eq:rotDyna_revised}
\end{equation}
where $\Omega$ is the strength of vision-based alignment and $D_{\rm R}$ is the rotational diffusion coefficient. The stochastic rotational noise ${\zeta}_i(t)$ is also Gaussian white noise with $\langle \zeta_i(t)\zeta_j(t')\rangle = \delta_{ij}\delta(t-t')$.

The first term on the right-hand side of Eq.~(\ref{eq:rotDyna_revised}) implements the vision-cone-based perception of the iABPs. In contrast to conventional active Brownian particles, each iABP senses only the particles lying within its finite vision cone (VC), characterized by a half-opening angle $\Theta_0$ and a vision range $R_{\rm v}$, as illustrated schematically in Fig.~\ref{fig:modelSchematic}. The normalization factor $N_{c,i}$ is defined as
\begin{equation}
N_{c,i}=
\sum_{j\in {\rm VC}}
e^{-r_{ij}/R_0},
\label{eq:neighborVC}
\end{equation}
where $R_0$ sets the characteristic decay length of visual sensitivity. Here, ${\bf \hat r}_{ij} = {({\bf r}_j-{\bf r}_i)}/{|{\bf r}_j-{\bf r}_i|} = (\cos\phi_{ij},\sin\phi_{ij})$, is the unit vector connecting particle $i$ to particle $j$, and $r_{ij}=|{\bf r}_j-{\bf r}_i|$ is their center-to-center separation. The summation in Eqs.~(\ref{eq:rotDyna_revised}) and (\ref{eq:neighborVC}) runs over all particles satisfying the geometric constraints
\begin{equation}
{\bf \hat r}_{ij}\cdot {\bf e}_i =
\cos(\phi_{ij}-\theta_i)
\geq
\cos\Theta_0
\quad \text{and} \quad
r_{ij}\le R_{\rm v}.
\end{equation}
Thus, only particles within the forward-vision cone and the finite sensing range contribute to the orientational dynamics. Throughout this work, we set $R_{\rm v}=4R_0$.

To investigate the effect of obstacles on the collective behavior of iABPs, we introduce a static circular obstacle of radius $R_{\rm obs}$ fixed at the center of the simulation box. Both particle-particle and particle-obstacle interactions are modeled through a shifted Lennard-Jones potential \cite{MidyaMacromol2025}
\begin{equation}
U(r)=
\begin{cases}
4\varepsilon
\left[
\left(
\dfrac{\sigma}{r-\Delta}
\right)^{12}
-
\left(
\dfrac{\sigma}{r-\Delta}
\right)^6
\right]
+\varepsilon,
& r\le r_c,\\[6pt]
0,
& r>r_c,
\end{cases}
\label{eq:sLJ}
\end{equation}
where $\varepsilon$ is the interaction strength and $\sigma$ is the diameter of an iABP. For particle-obstacle interactions, the shift parameter is chosen as $\Delta=R_{\rm obs}-\frac{\sigma}{2}$, with cutoff $r_c=\Delta+2^{1/6}\sigma$, ensuring purely repulsive excluded-volume interactions. For particle-particle interactions, we set $\Delta=0$ and $r_c=2^{1/6}\sigma$, which reduces the interaction to the standard Weeks-Chandler-Andersen (WCA) form.

Initially, a single obstacle is placed at the center of the simulation box, and the $N_{\rm act}$ particles are distributed randomly in the remaining accessible area while avoiding overlaps with each other and with the obstacle. The particle number is fixed by the packing fraction $\phi= {\pi \sigma^2 N_{\rm act}}/{4\left(L^2-N_{\rm obs}\pi R_{\rm obs}^2\right)}$, where $N_{\rm obs}~(=1)$ is the number of obstacles. For bulk systems, we set $N_{\rm obs}=0$. Periodic boundary conditions are imposed along both spatial directions. The equations of motion are integrated numerically using the Euler-Maruyama method \cite{Platen1999,Callegari2019} with a time step $\Delta t=10^{-4}\tau$. Length, energy, and time are measured in units of $\sigma$, $k_{\rm B}T$, and the microscopic time scale $\tau=\sqrt{m\sigma^2/(k_{\rm B}T)}$, respectively, where $m=1$ is introduced as a reference particle mass solely for defining the simulation time scale, as the overdamped equations of motion do not explicitly include inertial dynamics. 

Throughout this work, we set $\sigma=1$ and $k_{\rm B}T=1$. The translational diffusion coefficient is fixed at $D_{\rm T}=0.01\sigma^2/\tau$, corresponding to a Brownian diffusion time $\tau_{_{\rm B}}=\sigma^2/D_{\rm T}=100\tau$, while the rotational diffusion coefficient is chosen as $D_{\rm R}={8D_{\rm T}}/{\sigma^2}$.
The activity of the particles is characterized by the P\'eclet number $Pe={\sigma v_0}/{D_{\rm T}}$. To maintain approximately constant particle overlap during collisions over a wide range of activity, we choose the interaction strength as ${\varepsilon}/{k_{\rm B}T}=1+Pe$ \cite{RNegiSoftMatter2022}. Unless otherwise stated, all simulations are performed at fixed $\Omega/D_{\rm R}=80$ and $R_0=1.5\sigma$. Each run consists of $2\times10^8$ equilibration steps followed by $8\times10^8$ production steps, and all reported observables are averaged over at least ten independent realizations. In the limit $\Omega=0$, the model reduces to the standard ABP model, which serves as the reference system throughout this work.

\begin{figure}[tb]
    \centering
    \includegraphics[width=1.0\linewidth]{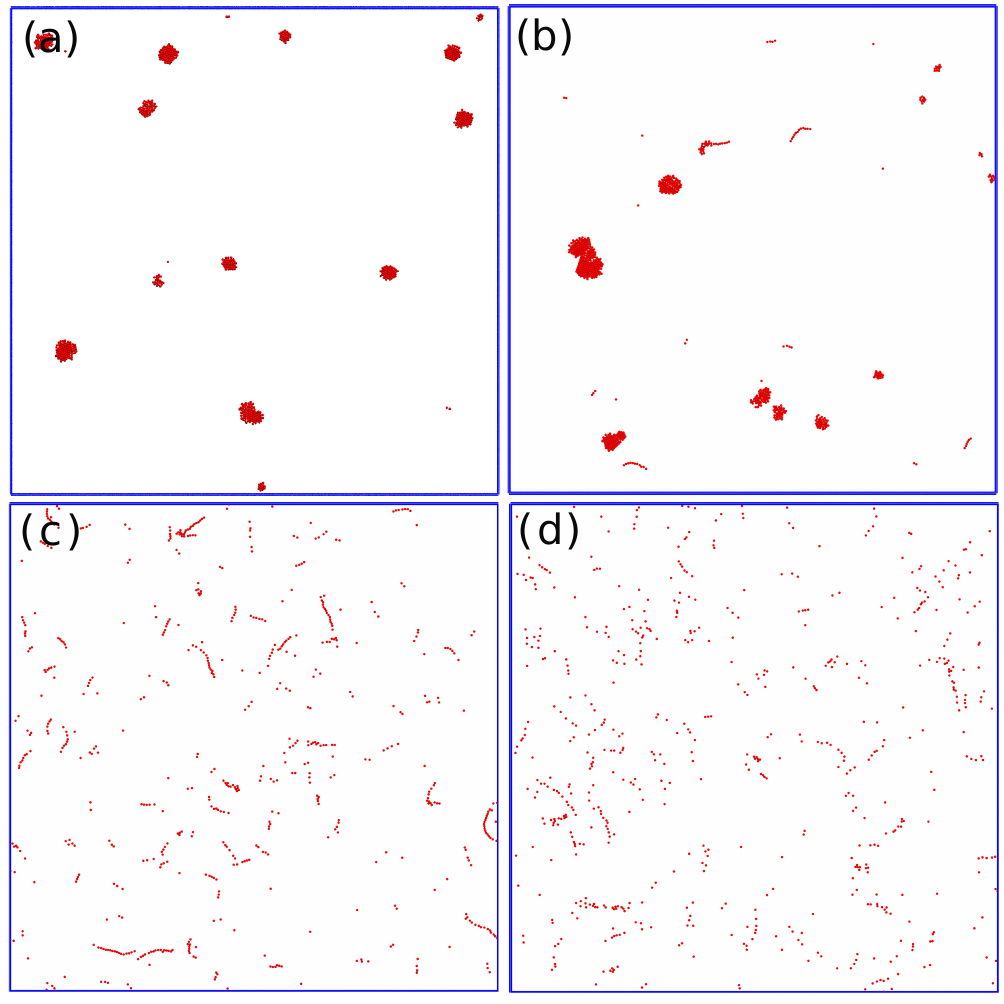}
    \caption{Representative snapshots for different P\'eclet numbers $Pe$ at fixed packing fraction $\phi \simeq 0.00785$, vision strength $\Omega/D_{\rm R}=80$, and vision angle $\Theta_0=\pi/6$ in a square box of size $L=250\sigma$, illustrating four distinct phases: (a) a dense compact aggregate at $Pe=70$, (b) a worm–aggregate coexistence phase at $Pe=100$, (c) worm-like chains at $Pe=500$, and (d) a nearly dilute phase at $Pe=1000$.}
    \label{fig:bulk_phases}
\end{figure}
\section{Results} \label{sec:results}
\subsection{Collective behavior of iABPs in bulk}
We perform simulations by randomly placing $N_{\rm act}=625$ iABPs in a square box of length $L=250\sigma$, corresponding to a packing fraction $\phi \simeq 0.0078$, and systematically vary the P\'{e}clet number $Pe$ and the vision cone angle $\Theta_0$. In Fig.~\ref{fig:bulk_phases} we present the steady-state morphologies of the system for different values of $Pe$ at fixed $\Theta_0=\pi/6$. At low activity ($Pe \lesssim 70$), the system initially forms small aggregated clusters [Fig.~\ref{fig:bulk_phases}(a)] that grow via collisions and coalescence, ultimately leading to a single macroscopic cluster in steady state. The aggregation process slows down due to an effective repulsion between clusters, arising from constituent particles that tend to orient toward the cluster center of mass (see Fig.~S1 in the Supplementary Information ~\cite{SI}), thereby hindering further merging of clusters. As $Pe$ increases, enhanced persistence in particle motion leads to the gradual elongation of compact clusters into worm-like structures, resulting in a worm--aggregate coexistence regime for $70 \lesssim Pe \lesssim 200$ and a worm phase for $200 \lesssim Pe \lesssim 1000$ [Fig.~\ref{fig:bulk_phases}(b,c)]. For larger activities ($Pe > 1000$), these worm structures become unstable, undergo fragmentation, and the system transitions to a dilute phase [Fig.~\ref{fig:bulk_phases}(d)].
    
To identify the different phases, we characterize the shape of the individual clusters using a shape anisotropy parameter $A$, defined as \cite{Arkin2013,Bera2023}
\begin{equation}
A = \frac{\lambda_{\rm max} - \lambda_{\rm min}}{\sqrt{\lambda_{\rm max}^2 + \lambda_{\rm min}^2}}.
\label{Eq:anisoParams}
\end{equation}
Here, $\lambda_{\rm max}$ and $\lambda_{\rm min}$ denote the largest and smallest principal moments of the gyration tensor of the cluster, respectively. The limiting values $A = 1$ and $A = 0$ correspond to perfectly rod-like and circular cluster shapes, respectively. In the steady state, $A$ is computed only for those clusters containing a minimum of five particles. Depending on the values of $Pe$ and $\Theta_0$, the distribution of the anisotropy parameter, $P(A)$, may exhibit either a single peak or a double-peak structure. A peak in $P(A)$ in the range $0 < A \lesssim 0.15$ identifies the compact aggregate phase, whereas a peak in the range $0.85 \lesssim A < 1$ corresponds to the worm-like phase. 

\begin{figure}[tb]
\centering
    \includegraphics[width=1.0\linewidth]{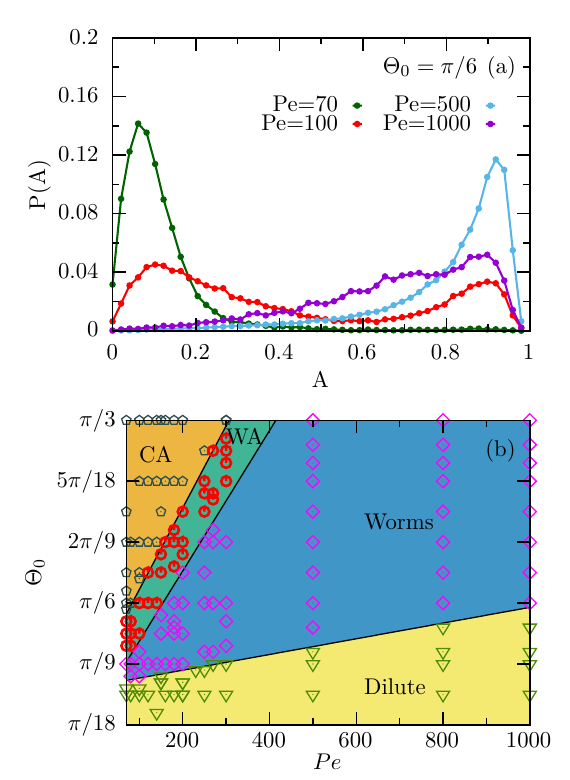}
    \caption{(a) The distribution of the anisotropy parameter $P(A)$ for different P\'{e}clet numbers $Pe=70, 100, 500,$ and $1000$ at a fixed vision angle $\Theta_0=\pi/6$, vision strength $\Omega/D_{\rm R}=80$ and packing fraction $\phi \simeq 0.00785$. 
    (b) Phase diagram in the $\Theta_0$–$Pe$ plane for the same choice of $\Omega/D_{\rm R}$ and $\phi$ as mentioned in (a). Four distinct phases, including the dilute phase, worm-like chains, worm–aggregate (WA) coexistence, and compact aggregates (CA), are presented in different colors. The black solid lines denote the coexistence boundaries and are intended as guides to the eye.}
    \label{fig:Asphericity_phDia}
\end{figure}

Figure~\ref{fig:Asphericity_phDia}(a) shows the distribution $P(A)$ for different values of $Pe$ at fixed $\Theta_0=\pi/6$. For $Pe=70$, $P(A)$ exhibits a single peak located at $A \approx 0.05$, indicating an aggregated-cluster phase. At $Pe=100$, the distribution becomes bimodal, with peaks at $A \approx 0.1$ and $A \approx 0.9$, signaling coexistence between compact aggregates and elongated worm-like chains. With further increase in $Pe$, the low-$A$ peak gradually diminishes, and $P(A)$ develops a single dominant peak at $A \approx 0.9$, consistent with a transition to a worm-like state. At even higher $Pe$, the peak broadens and shifts toward intermediate values, reflecting the destabilization of elongated structures and the onset of a dilute phase. The distribution function $P(A)$ exhibits similar behavior when the vision angle $\Theta_0$ is varied at fixed P\'{e}clet number $Pe=100$, see Fig.~2(a) in the supplementary information. 

Next, we construct the phase diagram in the $Pe$--$\Theta_0$ plane by analyzing the behavior of the distribution function $P(A)$. Within the considered range of $Pe$, four distinct phases are identified: compact aggregates (CA), worm--aggregate (WA) coexistence, pure worm phase, and dilute phase. These findings are consistent with the morphologies observed in the snapshots shown in Fig.~\ref{fig:bulk_phases}. These phases are summarized in Fig.~\ref{fig:Asphericity_phDia}(b) using different colors, while solid lines serve as guides to the eye and indicate approximate phase boundaries. Our results are broadly consistent with previous studies Ref.~\cite{RNegiSoftMatter2022}, although a systematic characterization of phase boundaries was not provided there. 

We find that at low $Pe$, the aggregate phase dominates over the other states. The extent of the vision-cone interval $\Delta \Theta_0$ corresponding to the worm–aggregate coexistence phase remains nearly constant over the explored range of $Pe$, whereas the $\Delta \Theta_0$ (separation between two boundaries) associated with the worm-like state increases with increasing $Pe$. In the limit $Pe \rightarrow 0$, extrapolation of the coexistence boundaries indicates that the system may undergoes a single structural transition from a dilute state to a compact aggregate state at a critical value of $\Theta_0$. This is expected because at low $Pe$, iABPs effectively behave as passive particles with attractive interactions at a finite value of $\Theta_0$. A similar analysis is performed to construct the phase diagram in the $\Omega$--$\Theta_0$ plane (see Fig.~2(b) in the Supplementary Information~\cite{SI}), which reveals trends consistent with Ref.~\cite{RNegiSoftMatter2022}.

\subsection{Collective behavior of iABPs in the presence of static circular obstacles}
\begin{figure*}[h!]
\centering
\includegraphics[width=1.0\linewidth]{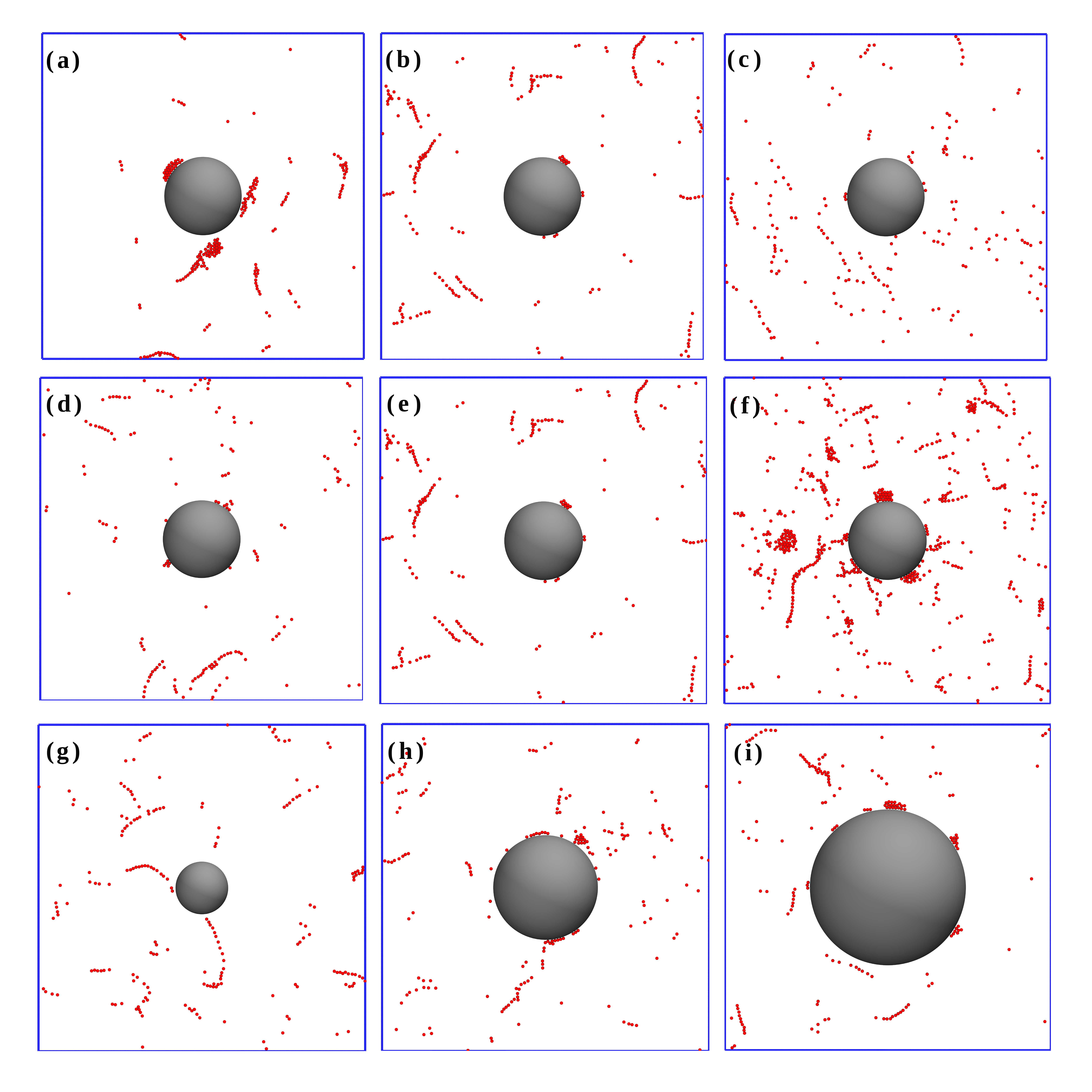}
\caption{Steady-state morphologies of iABPs (red) in the presence of a circular obstacle (gray) placed at the center of a simulation box of length $L=125\sigma$. Panels (a–c) correspond to peclet numbers $Pe=200, 500, 1000$ values at fixed obstacle size $R_{\rm obs}=15\sigma$ and packing fraction $\phi=0.01$; (d–f) show different packing fractions $\phi=0.00785, 0.01, 0.03$ at fixed $Pe=500$ and $R_{\rm obs}=15\sigma$; and (g–i) correspond to different obstacle radii $R_{\rm obs}= 10\sigma$, $20\sigma$, $30\sigma$ at fixed $\phi=0.01$ and $Pe=500$.}
\label{fig:snaps_withObs}
\end{figure*}

Next, we investigate the collective behavior of iABPs in the presence of an obstacle. Throughout this study, the vision angle is fixed at $\Theta_0=\pi/6$, while the P\'{e}clet number and packing fraction are varied over the ranges $200 \leq Pe \leq 1200$ and $0.005 < \phi < 0.03$, respectively, ensuring that the system remains in the worm-dominated phase. The effects of the P\'{e}clet number $Pe$, the packing fraction of active particles $\phi$, and the obstacle radius $R_{\rm obs}$ on the collective behavior are illustrated through a series of steady-state snapshots in Fig.~\ref{fig:snaps_withObs}. We find that iABPs start accumulating at the obstacle boundary for $R_{\rm obs} \gtrsim 10\sigma$, highlighting the role of curvature in boundary accumulation. At fixed $R_{\rm obs}=15\sigma$ and $\phi \simeq 0.01$, the degree of accumulation decreases with increasing $Pe$ [Fig.~\ref{fig:snaps_withObs}(a--c)], in contrast to conventional ABPs ($\Omega=0$) where accumulation increases with activity. At fixed $Pe$ and $R_{\rm obs}$, the degree of accumulation enhances with increasing $\phi$ due to enhancement of particles flux towards to the boundary; see Fig.~\ref{fig:snaps_withObs}(d--f). Moreover, larger obstacles promote aggregation by increasing particle–obstacle interactions; see Fig.~\ref{fig:snaps_withObs}(g--i). In contrast to conventional ABPs, which exhibit more homogeneous accumulation (see Figs.~S4-S6 in the Supplementary Information~\cite{SI}), iABPs show anisotropic aggregation, particularly at low $Pe$.

\begin{figure}[tb]
    \centering
    \includegraphics[width=1.0\linewidth]{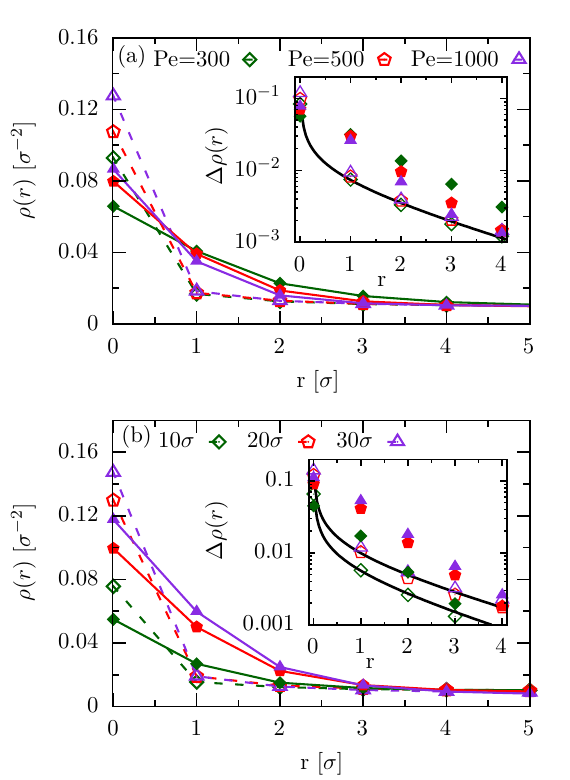}
    \caption{Radial density profile $\rho(r)$ for ABP (open symbols and dashed lines) and iABP (filled symbols and solid lines) systems at a fixed active-particle packing fraction $\phi \simeq 0.00785$. (a) $\rho(r)$ for different Péclet numbers $Pe$ at a fixed obstacle radius $R_{\rm obs}=15\sigma$. (b) $\rho(r)$ for different obstacle radii, $R_{\rm obs}=10\sigma$, $20\sigma$, and $30\sigma$, at a fixed Péclet number $Pe=500$. The insets show the excess density, $\Delta\rho(r)$, as a function of $r$ on a log-linear scale for the corresponding parameter sets. The solid black line represents the theoretical prediction given by Eq.~\ref{Eq:Radial_density}. 
    }
    \label{fig:radial_density}
\end{figure}

The distribution of particles around the obstacle is characterized by computing the radial density profile $\rho(r)$. For fixed $R_{\rm obs}=15\sigma$, the density profile $\rho(r)$ as a function of distance $r$ from the obstacle surface is shown in Fig.~\ref{fig:radial_density}(a) for different $Pe$, as indicated. The maximum value of $\rho(r\simeq 0.6\sigma)$ indicates the accumulation of particles at the obstacle boundary. For both ABPs and iABPs, $\rho(r \simeq 0.6\sigma)$ increases with increasing $Pe$, suggesting the increase of the first layer of particles at the boundary. At fixed $Pe$, however, $\rho(r\simeq 0.6\sigma)$ for iABPs is consistently lower than that for ABPs, indicating a reduced accumulation propensity of iABPs, arising from visual-perception-based interactions. At fixed $Pe=500$, the increase of obstacle radius $R_{\rm obs}$ enhances the surface accumulation, as evidenced by the value of $\rho(r\simeq 0.6\sigma$) in Fig.~\ref{fig:radial_density}(b). 

The excess radial density profile of dilute ABPs (without excluded-volume interactions) around a solid circular obstacle of radius $R_{\rm obs}$ can be expressed as \cite{PhysRevE.97.032606}
\begin{equation}
\Delta \rho(r) = \rho(r)-\rho_0 \sim \frac{\rho_0 v_0^2}{2D_R D_T}\sqrt{\frac{R_{\rm obs}}{r}}
\exp\left(-\frac{r}{\xi}\right),
\label{Eq:Radial_density}
\end{equation}
where $\rho_0$ is the bulk density, i.e., the particle density far from the obstacle surface, and $\xi$ is the characteristic decay length. The excess density $\Delta \rho(r)$ therefore exhibits an exponential decay, $\exp(-r/\xi)$, modulated by the curvature-induced algebraic prefactor $\sqrt{R_{\rm obs}/r}$ arising from the radial geometry of the system.

For conventional ABPs, the deposition of individual particles results in a homogeneous thin layer on the obstacle boundary. Consequently, $\Delta \rho(r)$ agrees well with the theoretical prediction of Eq.~(\ref{Eq:Radial_density}); see the insets of Fig.~\ref{fig:radial_density}(a,b). In contrast, iABPs exhibit a clear deviation from the predicted exponential decay because perception-mediated clustering leads to anisotropic aggregate formation along the obstacle boundary, particularly at low $Pe$. As $Pe$ increases, the system gradually approaches a dilute state in which cluster deposition is suppressed and single-particle deposition becomes increasingly dominant. Consequently, the density profiles of iABPs progressively approach those of conventional ABPs, see Fig.~\ref{fig:radial_density}(a,b).

\begin{figure}[tb]
\centering
\includegraphics[width=1.0\linewidth]{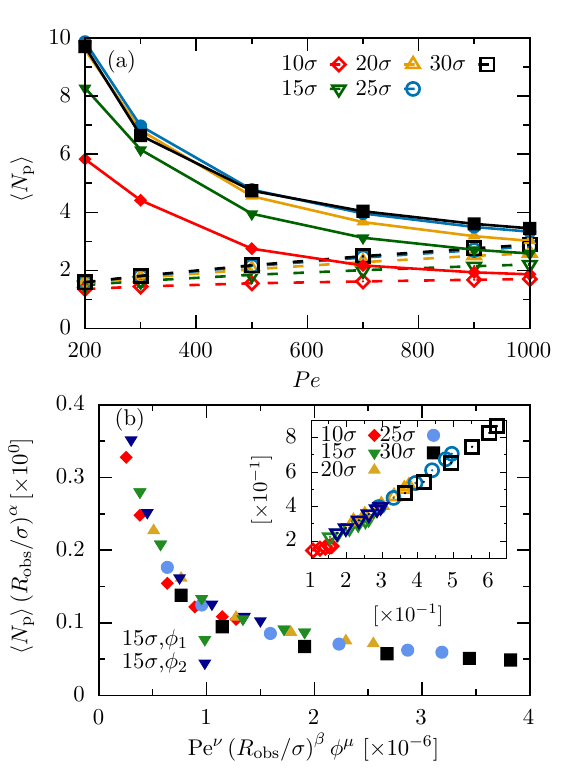}
\caption{(a) Average number of accumulated particles $\langle N_{\rm p} \rangle$ at the obstacle boundary as a function of the P\'{e}clet number $Pe$ for different obstacle radii $R_{\rm obs}$, as indicated. The results are shown for fixed $\Theta_0=\pi/6$, $\Omega/D_r=80$, and particle packing fraction $\phi \simeq 0.00785$. Dashed line with open symbols and solid lines with filled symbols correspond to ABPs and iABPs, respectively. (b) Data collapse of $\langle N_{\rm p} \rangle$ for different obstacle radii $R_{\rm obs}$ and particle packing fractions $\phi$, when plotted as $\langle N_{\rm p} \rangle (R_{\rm obs}/\sigma)^{\alpha}$ versus $Pe^{\nu}(R_{\rm obs}/\sigma)^{\beta}\phi^{\mu}$. For iABPs, the best data collapse is obtained with the exponents $\nu\simeq 1$, $\mu\simeq -1$, $\alpha\simeq -1.25$, and $\beta\simeq 1$. The inset shows the corresponding data collapse for ABPs, obtained with $\nu\simeq 1/3$, $\mu\simeq 0.5$, $\alpha\simeq 1$, and $\beta\simeq 1.25$.
}

\label{fig:Np_phi_c}
\end{figure}

\begin{figure}
    \centering
    \includegraphics[width=1.0\linewidth]{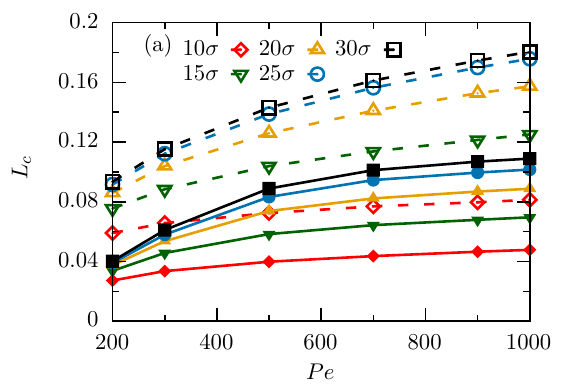}
    \caption{(a) fraction of particles, $L_{\rm c}$, around the obstacle as functions of the Péclet number $Pe$ for different obstacle radii $R_{\rm obs}$, as indicated. The results are shown at fixed $\Theta_0=\pi/6$, $\Omega/D_{\rm r}=80$, and packing fraction $\phi \simeq 0.00785$. Dashed and solid lines represent ABPs and iABPs, respectively.}
    \label{fig:LcvsPe}
\end{figure}

The average number of accumulated particles, $\langle N_{\rm p} \rangle$ at the obstacle boundary, and the boundary coverage fraction, $L_{\rm c} = \frac{\tilde{N}_{\rm act}\sigma}{2\pi R_{\rm obs}}$, where $\tilde{N}_{\rm act}$ is the instantaneous number of particles directly touching the obstacle boundary, as functions of $Pe$ are presented in Fig.~\ref{fig:Np_phi_c}(a) and Fig.~\ref{fig:LcvsPe}, respectively. At fixed obstacle radius $R_{\rm obs}$ and packing fraction $\phi$, both $\langle N_{\rm p} \rangle$ and $L_{\rm c}$ increase monotonically with increasing $Pe$ for conventional ABPs. In contrast, iABPs exhibit qualitatively opposite behavior to that of ABPs: $\langle N_{\rm p} \rangle$ decreases with increasing $Pe$. This behavior is closely linked to the distinct accumulation mechanisms. For conventional ABPs, the deposition of single particles at the obstacle boundary results in the formation of a homogeneous layer. As $Pe$ increases, more particles hit the obstacle and get trapped at the boundary, resulting in an increase in both $\langle N_{\rm p} \rangle$ and $L_{\rm c}$. 

For iABPs, the perception-mediated interactions promote the formation of worm-like chains in the bulk for a vision angle of $\Theta_0=\pi/6$, a particle packing fraction of $\phi \simeq 0.00785$, and P\'{e}clet numbers in the range $200 \lesssim Pe \lesssim 1200$; see the phase diagram in Fig.~\ref{fig:Asphericity_phDia}(b). As the P\'{e}clet number increases, the system gradually transitions to a dilute phase, leading to a reduction in both the average worm size and the total number of worm-like clusters; see Fig.~S3 in the Supplementary Information \cite{SI}. In the presence of an obstacle, these worm-like clusters frequently collide with the obstacle boundary. The leading particle has a high propensity to become trapped at the obstacle boundary because it moves largely independently of the particles trailing behind it and therefore behaves similarly to a conventional ABP.
Consequently, the remaining particles in the chain subsequently follow the leading particle and accumulate at the obstacle boundary.

The obstacle boundary serves as a nucleation site for the formation of compact aggregates. Upon collision, long worms (at low $Pe$) reorganize into compact aggregates mediated by interactions with the obstacle and subsequent collisions with isolated particles or other worms. As only a small fraction of the particles in these aggregates remains in direct contact with the boundary, the boundary coverage $L_{\rm c}$ remains low despite the large average number of accumulated particles $\langle N_{\rm p}\rangle$. The particles within the aggregates are highly mobile. Therefore, even a small local perturbation can transform the compact aggregates back into worm-like structures, allowing partial detachment from the boundary. Nevertheless, the continuous arrival of long worms replenishes the boundary layer, sustaining a large $\langle N_{\rm p}\rangle$. With increasing $Pe$, the system gradually enters the dilute phase, where long worms fragment into shorter worms and isolated particles become more abundant. These shorter worms are less likely to reorganize into compact aggregates and instead retain their elongated shape. As the worm size and the number of worms both decrease with increase $Pe$, leading to decrease in $\langle N_{\rm p}\rangle$, whereas the elongated morphology of the accumulated clusters leads to a monotonic increase in the boundary coverage, $L_{\rm c}$. The accumulation of particles is also strongly influenced by the obstacle radius, $R_{\rm obs}$, and the particle packing fraction, $\phi$. At fixed $Pe$, both $\langle N_{\rm p}\rangle$ and $L_{\rm c}$ increase with increasing $R_{\rm obs}$ and $\phi$ for both ABPs and iABPs; see Figs.~S7 and S8 in the Supplementary Information~\cite{SI}.

For iABPs, the data corresponding to different obstacle radii $R_{\rm obs}$ and particle packing fractions $\phi$ exhibit data collapse and form a single master curve when plotted as $\langle N_{\rm p} \rangle (R_{\rm obs}/\sigma)^{\alpha}$ versus $Pe^{\nu}(R_{\rm obs}/\sigma)^{\beta}\phi^{\mu}$, with the correct choice of exponents $\alpha \simeq -1.25$, $\beta \simeq 1$, $\nu \simeq 1$, and $\mu \simeq -1$; see Fig.~\ref{fig:Np_phi_c}(b). In contrast, for ABPs, a reasonably good data collapse is obtained with the optimal exponents $\alpha \simeq 1$, $\beta \simeq 1.25$, $\nu \simeq 0.33$, and $\mu \simeq 0.5$, see the inset of Fig.~\ref{fig:Np_phi_c}(b). The iABP master curve exhibits a power-law decay with an exponent close to $-1$, whereas the ABP master curve is approximately linear, implying the simplified scaling relations
\begin{equation}
\langle N_{\rm p} \rangle \sim
\begin{cases}
Pe^{-1}\left(\dfrac{R_{\rm obs}}{\sigma}\right)^{0.25}\phi, & \text{iABPs},\\[6pt]
Pe^{0.33}\left(\dfrac{R_{\rm obs}}{\sigma}\right)^{0.25}\phi^{0.5}, & \text{ABPs}.
\end{cases}
\label{Eq:scaling}
\end{equation}
The observed data collapse reveals a universal law in which the effects of particle activity, obstacle size, and particle packing fraction combine into a single scaling variable which effectively control the particle accumulation. The opposite dependence on $Pe$ highlights the distinct accumulation mechanisms in the two systems. In iABPs, increasing activity suppresses particle accumulation owing to the activity-induced transition from long worm-like state to a dilute state, together with perception-mediated escape from the obstacle. In contrast, increasing activity enhances particle accumulation in ABPs through the increased flux of active particles toward the boundary. In both systems, particle accumulation depends only weakly on the obstacle size $\langle N_{\rm p} \rangle \sim (R_{\rm obs}/\sigma)^{0.25}$, while the different $\phi$ dependence reflect the contrasting roles of bulk particle supply and collective interactions in governing particle accumulation. 

\begin{figure}[tb]
\centering
\includegraphics[width=\linewidth]{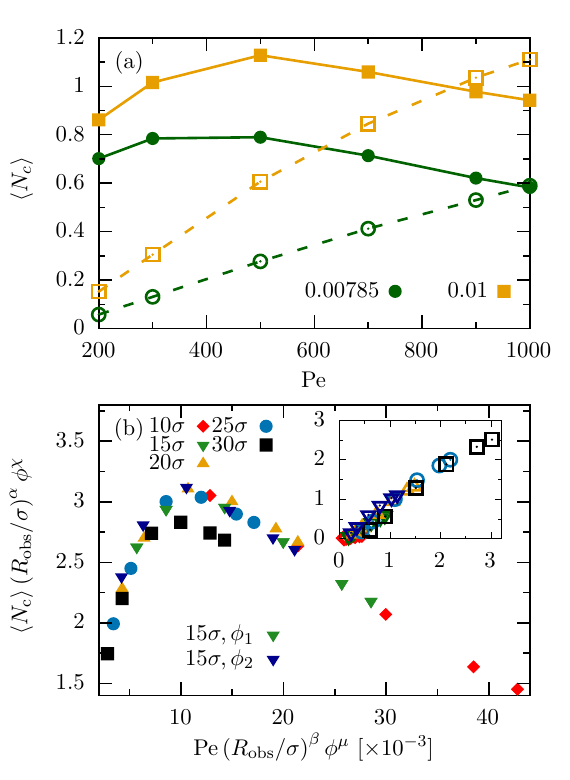}
\caption{(a) Average number of clusters, $\langle N_{\rm c} \rangle$, at the obstacle boundary as a function of the P\'{e}clet number, $Pe$, for different packing fractions, $\phi$, at a fixed obstacle radius $R_{\rm obs}=15\sigma$. Dashed and solid lines correspond to ABPs and iABPs, respectively. (b) Scaling of $\langle N_{\rm c} \rangle$ with $Pe$ for different obstacle radii, $R_{\rm obs}$, as indicated. For iABPs, the data for different obstacle radii collapse onto a single master curve when plotted as $\langle N_{\rm c} \rangle (R_{\rm obs}/\sigma)^{\alpha} \phi^{\chi}$ {\it vs} $Pe\,(R_{\rm obs}/\sigma)^{\beta} \phi^{\mu}$, with scaling exponents $\alpha \approx -1.75$, $\beta \approx -1$, and $\chi = \mu \simeq -1.25$. The inset shows the corresponding scaling for ABPs, yielding a master curve with $\alpha \approx 0$ and $\beta \approx 1.75$, $\mu \approx 1$ and $\chi=0$.}
\label{fig:cluster_dia}
\end{figure}

In Fig.~\ref{fig:cluster_dia}(a), we show the average number of clusters $\langle N_{\rm c} \rangle$ containing at least five particles and accumulated at the obstacle boundary as a function of $Pe$ for different packing fractions $\phi$ and at a fixed obstacle radius $R_{\rm obs}=15\sigma$. For ABPs, $\langle N_{\rm c} \rangle$ increases monotonically with $Pe$, indicating the formation of an increasing number of isolated clusters at the obstacle boundary as $\langle N_{\rm p} \rangle$ increases with activity. In contrast, iABPs exhibit a non-monotonic dependence on $Pe$: $\langle N_{\rm c} \rangle$ initially increases, reaches a maximum at intermediate $Pe$, and then decreases at higher $Pe$. This behavior arises from the competition between the increasing collision frequency of worm-like clusters with the obstacle and the simultaneous reduction in both the size and the number of worm-like clusters as the system gradually transitions from the worm phase to the dilute phase with increasing activity. The average number of clusters $\langle N_{\rm c} \rangle$ also increases with the obstacle radius $R_{\rm obs}$, because the average number of accumulated particles $\langle N_{\rm p} \rangle$ increases sublinearly with obstacle size; see Fig.~S7 in the Supplementary Information~\cite{SI}.

For iABPs, the $\langle N_{\rm c} \rangle$ data for different obstacle radii $R_{\rm obs}$ and particle packing fractions $\phi$ collapse onto a single master curve when plotted $\langle N_{\rm c} \rangle (R_{\rm obs}/\sigma)^{\alpha}\phi^{\chi}$ versus $Pe\,(R_{\rm obs}/\sigma)^{\beta}\phi^{\mu}$ for the choice of the exponents $\alpha \simeq -1.75$, $\beta \simeq -1$, and $\chi=\mu \simeq -1.25$; see Fig.~\ref{fig:cluster_dia}(b). The master curve also exhibits a nonmonotonic behavior, leading to a peak at intermediate activity. Although the packing-fraction exponent is identical on both axes, it cannot be eliminated because $\phi$ rescales the effective activity and shifts the peak position. In contrast, for ABPs, the data collapse onto a nearly linear master curve when $\langle N_{\rm p} \rangle$ plotted against the combined scaling variable $Pe(R_{\rm obs}/\sigma)^{1.75}\phi$ (here we choose $\alpha \simeq 0$, $\beta \simeq 1.75$, $\chi \simeq 0$ and $\mu \simeq 1$); see inset of Fig.~\ref{fig:cluster_dia}(b). This collapse indicates that particle accumulation is controlled by a single scaling variable, with activity enhancing particle flux toward the obstacle, larger obstacles providing a larger capture region, and higher packing fractions increasing the bulk particle supply.

\begin{figure}[tb]
\centering
\includegraphics[width=\linewidth]{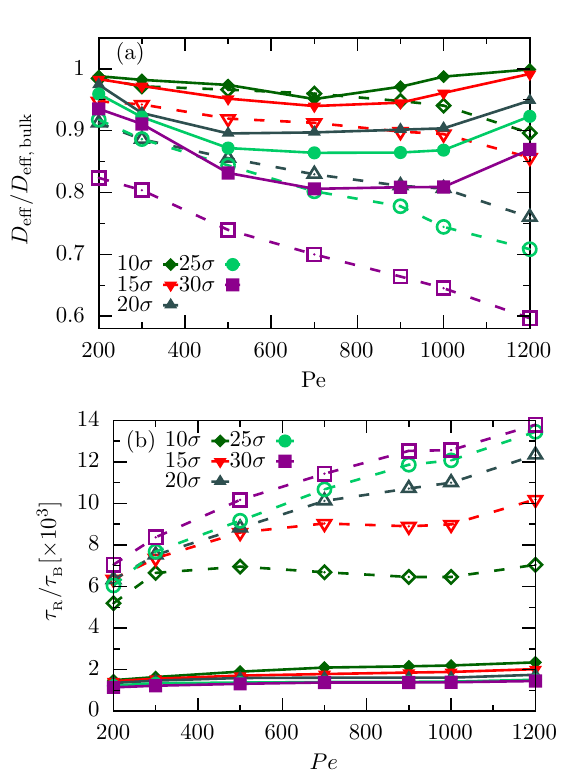}
\caption{(a) Normalized effective diffusion coefficient $D_{\rm eff}/D_{\rm {eff,\, bulk}}$  and (b) normalized average residence time $\tau_{_{\rm R}}/\tau_{_{\rm B}}$, as functions of the P\'{e}clet number $Pe$ for different obstacle radii $R_{\rm obs}$ and at a fixed packing fraction $\phi \simeq 0.00785$. Solid lines with filled symbols correspond to the iABPs, while dashed lines with open symbols correspond to the ABPs. 
}
\label{fig:Diif_ResTime}
\end{figure}

To study the dynamics of the particles we compute the mean-squared displacement (MSD), defined as
\begin{equation}
\langle \delta r^2(t)\rangle=\frac{1}{N_{\rm act}} \sum_{i=1}^{N_{\rm act}} \left\langle \left({\bf r}_i(t)-{\bf r}_i(0)\right)^2 \right\rangle.
\label{EqMSD}
\end{equation}
At early times, the particles exhibit ballistic motion characterized by $\langle \delta r^2(t)\rangle \sim t^2$, while at longer times they cross over to diffusive behavior with $\langle \delta r^2(t)\rangle \sim t$, as shown in Fig.~S9 of the Supplementary Information ~\cite{SI}. From the long-time diffusive regime, we extract the effective diffusion coefficient $D_{\rm eff}$ and observe that it increases with increasing $Pe$ for both iABPs and ABPs, reflecting the enhanced propulsion at higher activity.

To quantify the effect of obstacle-induced accumulation on the overall dynamics, we evaluate the normalized effective diffusion coefficient  $D_{\rm eff}/D_{\rm {eff,\, bulk}}$, where $D_{\rm eff,\,bulk}$ denotes the effective diffusion coefficient of the corresponding bulk system (i.e., in the absence of obstacles). For iABPs,  $D_{\rm eff}/D_{\rm {eff,\, bulk}} \approx 1$ at low $Pe$, as the system forms slowly moving large worms-like chains that interact weakly with the obstacle, resulting in negligible differences between the bulk and confined dynamics. As $Pe$ increases, these long worms fragment into smaller fast moving worms, which interact more frequently with obstacle and accumulate at the boundary. As a result the relative diffusion decrease, leading to a suppression of  $D_{\rm eff}/D_{\rm {eff,\, bulk}}$. At higher $Pe$, $\langle N_{\rm p}\rangle$ decreases as particles increasingly escape from the obstacle boundary and redistribute into the bulk, resulting in a recovery of $D_{\rm eff}/D_{\rm {eff,\, bulk}} \simeq 1$. In contrast, for ABPs, increasing $Pe$ promotes enhanced single-particle accumulation near the obstacle, which leads to a monotonic suppression of  $D_{\rm eff}/D_{\rm {eff,\, bulk}}$. Furthermore, at fixed $Pe$ and $\phi$, increasing the obstacle size $R_{\rm obs}$ enhances particle accumulation and thereby further suppresses the relative diffusion for both systems. The qualitative behavior of  $D_{\rm eff}/D_{\rm {eff,\, bulk}}$ remains unchanged with varying packing fraction $\phi$ at fixed obstacle size, as shown in  Fig.~S9(b) of the Supplementary Information ~\cite{SI}. 

Finally, we compute the average residence time, $\tau_{_{\rm R}}$, which quantifies the average time individual particles spend near the obstacle boundary. The boundary region is defined by $R_{\rm obs} < r < R_{\rm obs} + \delta$, where $\delta \simeq 0.1\sigma \ll R_{\rm obs}$. For ABPs (dashed lines), the residence time increases with increasing $Pe$, particularly for larger obstacle sizes, as shown in Fig.~\ref{fig:Diif_ResTime}(b). At higher activity, ABPs undergo persistent collisions with the hard boundary and remain trapped near the obstacle surface. As $Pe$ increases, the average radial orientation toward the obstacle center $\langle \cos\theta \rangle$, where $\cos\theta= \mathbf{e}_i\cdot (\mathbf{r}_i-\mathbf{r}_{\rm{obs}})/|\mathbf{r}_i-\mathbf{r}_{\rm{obs}}|$, $\mathbf{r}_{\rm{obs}}$ being the obstacle location, increases, whereas the average tangential orientation $\langle \sin\theta \rangle$ decreases (see Fig.~S10 in the Supplementary Information~\cite{SI}). This suppresses motion along the boundary, causing particles to remain near the obstacle for longer times and thereby increasing $\tau_{_{\rm R}}$. The effect is further enhanced for larger obstacles because their lower curvature prolongs particle--boundary interactions.

In contrast, the residence times of iABPs are orders of magnitude smaller than those of the corresponding ABP systems, indicating that visual-perception interactions facilitate efficient escape from the boundary region. At a fixed obstacle radius, $R_{\rm obs}$, $\tau_{_{\rm R}}$ exhibits only a weak dependence on $Pe$, increasing slightly as the average radial orientation toward the obstacle center $\langle \cos\theta \rangle$ increases; see Fig.~S10 in the Supplementary Information~\cite{SI}. At fixed $Pe$, however, $\tau_{_{\rm R}}$ decreases with increasing obstacle radius, opposite to the trend observed for conventional ABPs. This behavior arises because the leading particle of a worm, which behaves similarly to a conventional ABP, moves tangentially along the obstacle surface owing to its slow orientational diffusion, while the remaining particles follow collectively. As the obstacle radius increases, the worms become increasingly aligned with the boundary, resulting in a dominant tangential orientation $\langle \sin\theta \rangle$ and a corresponding suppression of the radial orientation $\langle \cos\theta \rangle$, see Fig.~S10 in the Supplementary Information~\cite{SI}. Consequently, the worms slide more efficiently along the obstacle surface and detach more readily, leading to a reduction in the average residence time $\tau_{_{\rm R}}$.

\section{Conclusion} \label{sec:conclusion}
In conclusion, we have investigated the collective behavior of two-dimensional active Brownian particles with excluded-volume interactions, both in bulk and in the presence of a static circular obstacles, incorporating perception-mediated interactions through a vision-based steering mechanism. In bulk, these intelligent active Brownian particles exhibit four different collective states: compact aggregates, worm-like structures, worm–aggregate coexistence states, and dilute gas-like phases. The transitions among these states governed by the interplay of activity, vision angle, and interaction range. To characterize these morphologies, we introduce a cluster anisotropy parameter $A$, constructed in terms of the eigenvalues of the gyration tensor of the clusters. The distribution of $A$ exhibits a single peak for pure aggregate and worm-like phases, while a bimodal distribution is observed in the worm–aggregate coexistence states. This parameter thus enables robust phase identification and the construction of the corresponding phase diagrams.

The introduction of an circular static obstacle alters the collective behavior of active particles. While conventional ABPs accumulate isotropically and exhibit enhanced trapping with increasing activity, iABPs display anisotropic aggregation and reduced boundary accumulation due to perception-driven reorientation. As a result, the average residence time $\langle \tau \rangle$ of iABPs near the obstacle is reduced by orders of magnitude compared to ABPs, indicating a strong suppression of boundary trapping. Consequently, the transport behavior of iABPs differ qualitatively: in contrast to ABPs, for which the normalized effective diffusion coefficient decreases monotonically with increasing $Pe$, iABPs exhibit a nonmonotonic dependence on $Pe$. This behavior is closely linked to the underlying morphological transitions of the system, evolving from compact aggregates to worm–aggregate coexistence, worm-like structures, and eventually a dilute phase. Moreover, the reduced affinity of iABPs for curved boundaries becomes more pronounced with increasing obstacle size, highlighting the important role of geometry in shaping active transport.

These findings demonstrate that perception-mediated interactions provide a powerful mechanism to control both the spatial organization and transport properties of active matter in crowded environments. Such control may be exploited in the design of adaptive synthetic active systems, including targeted delivery platforms and programmable microswimmers, where navigation through complex geometries is essential. Furthermore, our results offer insights into biological systems in which agents rely on local sensing and decision-making to navigate crowded environments. Future studies could explore the interplay of perception with hydrodynamic interactions, complex obstacle geometries, and collective decision-making strategies, thereby advancing the understanding of intelligent active matter in realistic settings.
\\
\section*{Author contributions} 
JM conceived and designed the research. SNS, AB, and JM developed the simulation code. SNS and JM carried out the simulations and analyzed the data. SNS, AB, and JM wrote the manuscript.

\section*{Data Availability Statement}
Data will be made available upon request.

\section*{Conflicts of interest} 
The authors declare that there is no conflict of interest.

\section*{Acknowledgment}
JM thanks Gerhard Gompper (Forschungszentrum J\"{u}lich, Germany) and A. V. Anil Kumar (NISER, India) for insightful discussions and valuable suggestions. JM and SNS acknowledge financial support and research facilities provided by IIT Bhubaneswar.

\bibliography{reference}
\end{document}